\documentstyle[12pt]{article}

\newcommand{\mysection}{\setcounter{equation}{0}\section}

\begin{document}
\hfill{ITP-SB-93-1}
\vskip 0.1cm
\hfill{FERMILAB-Pub-93/033-T}
\vskip 0.5cm
\centerline{\large\bf { $O(\alpha_S)$ Corrections to the
Photon Structure Functions}}
\vskip 0.2cm
\centerline{\large\bf{ $F^\gamma_2(x,Q^2)$ and $F_L^\gamma(x,Q^2)$}}
\vskip 0.4cm
\centerline{\sc E. Laenen}
\vskip 0.3cm
\centerline{\it Fermi National Accelerator Laboratory,}
\centerline{\it P.O. Box 500, MS 106}
\centerline{\it Batavia, Illinois 60510}
\vskip 0.3cm
\centerline {\sc S. Riemersma, J. Smith }
\vskip 0.3cm
\centerline{\it Institute for Theoretical Physics,}
\centerline{\it State University of New York at Stony Brook,}
\centerline{\it Stony Brook, New York 11794-3840}
\vskip 0.3cm
\centerline{and}
\vskip 0.3cm
\centerline{\sc W. L. van Neerven}
\vskip 0.3cm
\centerline{\it Instituut Lorentz,}
\centerline{\it University of Leiden,}
\centerline{\it P.O.B. 9506, 2300 RA, Leiden,}
\centerline{\it The Netherlands.}
\vskip 0.3cm
\centerline{February 1993}
\vskip 0.3cm
\centerline{\bf Abstract}
\vskip 0.4cm

We examine the QCD corrections to the structure
functions $F_2^\gamma(x,Q^2)$ and $F_L^\gamma(x,Q^2)$
for a real photon target.
The pointlike photon contributions from light and heavy quarks
are computed through $O(\alpha\alpha_S)$. A parameterization of
the hadronic (resolved) photon contribution is also included
and a comparison is made with the present experimental data.
We find that while the pointlike contributions are large
for $F_2^\gamma(x,Q^2)$, they are the dominant part of $F_L^\gamma(x,Q^2)$.
For $Q^2 > 50$ $({\rm GeV}/c)^2$ the charm component is
as least as large as the light quark component.

\vfill
\newpage
\mysection{Introduction}
In the past two decades there has been considerable interest
in the study of photon-photon interactions in electron-positron colliders.
When one photon is virtual and the other one is almost real
the analogy with deep-inelastic electron-nucleon scattering motivated
the introduction of the corresponding structure functions
$F_k^\gamma(x,Q^2)$ $(k=2,L)$ for the photon.
It was originally observed by Witten \cite{ew} that $\underline{\rm both}$
the $x$ and $Q^2$ dependence of these structure functions
were calculable in perturbative QCD (pQCD) at asymptotically
large $Q^2$. Thus from a theoretical point of view
this process should provide a much more thorough test of
pQCD than the corresponding deep-inelastic scattering off a
nucleon target, where only the $Q^2$ evolution
of the structure functions is calculable. The original
optimism subsided once it was realized that there were complications with
experimental confirmation of this prediction
at experimental (non-asymptotic) values of $Q^2$
\cite{bill}\,,\,\cite{gr}.  For recent reviews see \cite{jf}.
In particular at small $Q^2$ there is a contamination of the purely
pointlike (unresolved) pQCD contribution by the hadronic (resolved) component
of the photon. This latter piece is not calculable
in pQCD and must be extracted from experimental data.
One of the approaches used is to describe this hadronic piece by parton
densities in the photon. For parameterizations see
\cite{dg}, \cite{acl}, \cite{grv} and \cite{gs}.
For a different approach see \cite{fkp}.

In this paper we will investigate higher-order pQCD corrections
in the deep-inelastic structure functions containing
both light (massless) quarks and heavy (massive) quarks.
We report the results of including these pQCD corrections
through order $\alpha\alpha_S$.
These corrections have not been considered earlier in the literature.
They are the Abelian analogues of the order $\alpha_S^2$
corrections to light-and heavy-quark production
in deep-inelastic lepton hadron scattering
contributing to the structure functions $F_k(x,Q^2)$ $(k=2,L)$.
For light quarks the non-Abelian results were reported in \cite{zn}
while for heavy quarks they were reported in \cite{lrsn1}.

The deep-inelastic structure function $F_2^\gamma(x,Q^2)$
was originally measured by the PLUTO collaboration
\cite{pluto} at PETRA using single tag events in the
reaction $e^- + e^+ \rightarrow e^- + e ^+ + {\rm hadrons}$.
In the past few years there have been a series of new measurements
at PETRA, PEP and TRISTAN by several groups,including CELLO \cite{cello},
TPC2$\gamma$ \cite{tpc},TASSO \cite{tasso}, JADE \cite{jade},
AMY \cite{amy}, VENUS \cite{venus} and TOPAZ \cite{topaz}.
All these groups concentrated on the measurement of the light-quark
contribution to $F_2^\gamma(x,Q^2)$.
The heavy-quark component (mainly charm)
has been hard to extract due to problems identifying
charmed particle decays.
In the near future higher-luminosity runs at TRISTAN
should yield some information on heavy-quark (mainly charm)
production and this is one reason that we study it here.
We should mention that there was a
previous investigation of pQCD corrections to heavy quark
production in \cite{hr}, where it was assumed that both photons
were off-mass-shell and a small value for the
photon virtuality was chosen for generating numerical results. Since
these authors did not therefore encounter mass singularities they
had no need to perform any mass factorization. Hence their
method was different from the one we adopt.
Finally there exists a possibility that the longitudinal structure function
$F_L^\gamma(x,Q^2)$ can be measured at LEP \cite{ali}.
Therefore we will also present the higher-order pQCD corrections to
$F^\gamma_L(x,Q^2)$ which have not been reported previously.

Two-photon reactions are important to understand as
background processes to the normal $s$-channel reactions at present
and future $e^- e^+$ colliders. The latter machines will
have a large amount of beamstrahlung \cite{dg1}, \cite{eghn}.
Therefore a basic input is the parton density
in a photon, which is one of the topics we discuss.

The paper is organized as follows. In section 2 we
present the pQCD corrections up to order $\alpha\alpha_s$ which are used
in our calculations. In section 3 we discuss
the effects of the higher-order corrections to the
structure functions $F_k^\gamma(x,Q^2)$ $(k=2,L)$ including
the $O(\alpha\alpha_s)$ contributions from the light and heavy
quarks.
\newpage
\mysection{Higher-Order Corrections to the Photon Structure Functions}
The deep-inelastic photon structure functions denoted by
$F^\gamma_k(x,Q^2)$ $(k=2,L)$ are measured in $e^- e^+ $ collisions
via the process (see fig.1)
\begin{equation}
e^-(p_e) + e^+ \rightarrow e^-(p_e') + e^+ + X\,,
\end{equation}
where $X$ denotes any hadronic state which is allowed by quantum-number
conservation laws. When the outgoing electron is tagged then
the above reaction is dominated by the photon-photon collision reaction
(see fig.1)
\begin{eqnarray}
\gamma^*(q) + \gamma(k) \rightarrow X\,,
\end{eqnarray}
where one of the photons is highly virtual and the other one
is almost on-mass-shell. The process (2.1) is described by the
cross section
\begin{eqnarray}
\frac{d^2\sigma}{dxdy} &=& \int dz \, z\, f_\gamma^e (z, \frac{S}{m_e^2})
\,  \frac{2\pi\alpha^2 S}{Q^4} \nonumber \\&&
\left[ \{ 1 + (1-y)^2\} F^\gamma_2(x,Q^2) -y^2 F^\gamma_L(x,Q^2) \right]\,,
\end{eqnarray}
where $F^\gamma_k(x,Q^2)$ $(k=2,L)$ denote the deep-inelastic
photon structure functions and $\alpha = e^2/4\pi$
is the fine structure constant.
Furthermore the off-mass-shell photon and the on-mass-shell photon
are indicated by the four-momenta $q$ and $k$
respectively with $q^2 = -Q^2 <0$
and $k^2 \approx 0$. Because the photon with momentum $k$ is
almost on-mass-shell, expression (2.3) is written in the Weizs\"acker-Williams
approximation. In this approximation the function $f^e_\gamma(z,S/m_e^2)$
is the probability of finding a photon $\gamma(k)$ in the positron,
(see fig.1). The fraction of the energy of the positron
carried off by the photon is denoted by $z$ while
$\sqrt{S}$ is the c.m. energy of the electron-positron system.
The function $f^e_\gamma(z,S/m_e^2)$ is given by (see \cite{sn})
\begin{equation}
f^e_\gamma(z,\frac{S}{m_e^2}) = \frac{\alpha}{2\pi}
\frac{1 + (1-z)^2}{z} \ln \frac{(1-z)(zS-4m^2)}{z^2m_e^2}\,,
\end{equation}
provided a heavy quark with mass $m$ is produced.
The scaling variables $x$ and $y$ are
defined by
\begin{equation}
x = \frac{Q^2}{2k\cdot q} \,, \qquad y = \frac{k\cdot q}{k \cdot p_e}
\,, \qquad q = p_e - p_e'\,,
\end{equation}
where $p_e$, $p_e'$ are the momenta of the incoming and outgoing
electron respectively.
Following the procedure in \cite{gr1} the photon structure functions
in the QCD-improved parton model have the following form
\begin{eqnarray}
&& \frac{1}{\alpha} F_k^\gamma(x,Q^2) =
x \int_x^1 \frac{dz}{z} \, \left[ \left(\frac{1}{n_f}
\sum_{i=1}^{n_f} e_i^2 \right)\{\Sigma^\gamma(\frac{x}{z},M^2)
\,{\cal C}_{k,q}^S(z,\frac{Q^2}{M^2}) \right. \nonumber \\ &&
\qquad \left. +  g^\gamma(\frac{x}{z} ,M^2)
\,{\cal C}_{k,g}(z,\frac{Q^2}{M^2})\}
+\Delta^\gamma(\frac{x}{z},M^2) \,{\cal C}_{k,q}^{NS}(z,\frac{Q^2}{M^2})
\right]\nonumber \\ &&
\qquad +\frac{3}{4\pi} x \left[\left(\sum_{i=1}^{n_f} e^4_i\right)
\,{\cal C}_{k, \gamma}(x,\frac{Q^2}{M^2})
+ e_H^4\,{\cal C}_{k,\gamma}^H (x,Q^2,m^2)\right]
 \,.
\end{eqnarray}
Here $\Sigma^\gamma$ and $\Delta^\gamma$ represent the singlet
and non-singlet combinations of the parton densities in the photon
respectively while the gluon density is represented by $g^\gamma$.
The same notation also holds for the hadronic (Wilson) coefficient
functions ${\cal C}_{k,i}$ ($i=q,g$) where
${\cal C}_{k,q}^S$ and  ${\cal C}_{k,q}^{NS}$ stand
for the singlet and non-singlet coefficient functions respectively,
and ${\cal C}_{k,g}$ denotes the gluonic coefficient function.
The photonic coefficient functions for massless and massive
quarks are given by ${\cal C}_{k,\gamma}$ and
${\cal C}_{k,\gamma}^H$ respectively,
where $m$ in (2.6) denotes the heavy-quark mass.
The index $i$ in (2.6)
runs over all light flavors provided they can be produced in the final
state ($n_f$ is the number of light flavors) and $e_i$, $e_H$ stand for
the charges of the light and heavy quarks respectively in
units of $e$.
The parton densities as well as the coefficient functions depend
on the mass factorization scale $M$ except for the ${\cal C}^H_{k,\gamma}$
which can be calculated in pQCD without performing mass factorization.
Notice that in addition to the mass factorization scale $M$ the
quantities in (2.6) also depend on the renormalization scale
$\mu$ which appears in the pQCD corrections via $\alpha_s(\mu^2)$.
However in this paper we will put $\mu$ = $M$.
Because of the origin of the photonic parton densities
and the two different types of coefficient functions (photonic and
hadronic) we will call the first term, represented by the integral, the
$\underline{\rm hadronic}$ (resolved) photon part, and the second term
the $\underline{\rm pointlike}$ (unresolved) photon part. The latter
can be split into a light-quark contribution
due to ${\cal C}_{k,\gamma}$ and a heavy-quark contribution due to
${\cal C}_{k,\gamma}^H$.

In the subsequent discussions in this paper we will neglect all pQCD
corrections beyond the first order in $\alpha_s$ so that we can put
${\cal C}^S_{k,q} = \,{\cal C}^{NS}_{k,q} = \,{\cal C}_{k,q}$.
In this case (2.6) can be simplified as follows
\begin{eqnarray}
&& \frac{1}{\alpha} F_k^\gamma(x,Q^2) =
x \int_x^1 \frac{dz}{z} \, \left[
Q^\gamma(\frac{x}{z},M^2)
\,{\cal C}_{k,q}(z,\frac{Q^2}{M^2}) \right. \nonumber \\ && \left.
\qquad\quad + \left(\frac{1}{n_f}\sum_{i=1}^{n_f}e_i^2 \right)
g^\gamma(\frac{x}{z},M^2)
\,{\cal C}_{k,g}(z,\frac{Q^2}{M^2})  \right] \nonumber \\ &&
\qquad + \frac{3}{4\pi} x \left[\,\left(\sum_{i=1}^{n_f} e^4_i\right)
\,{\cal C}_{k,\gamma}(x,\frac{Q^2}{M^2})
+ e_H^4 \,{\cal C}_{k,\gamma}^H(x,Q^2,m^2)\right] \,,
\end{eqnarray}
with
\begin{eqnarray}
Q^\gamma(z,M^2) &=& \,\left(\frac{1}{n_f}\sum_{i=1}^{n_f} e_i^2
\right) \Sigma^\gamma(z,M^2)
+ \Delta^\gamma(z,M^2) \nonumber \\ &&
= 2 \sum_{i=1}^{n_f} e_i^2 q_i^\gamma(z,M^2)\,,
\end{eqnarray}
where we have set $q_i^\gamma= \bar q_i^\gamma$ in the above equations.
{}From now on we will only use (2.7) for our calculations, the results
of which will be presented in the plots in Section 3.

Starting with the parton densities we will follow the prescription
in \cite{dg} (where $\Sigma^\gamma$ has
the same meaning and $\Delta^\gamma = q^\gamma_{NS}$)
and, in the case where all quarks are light, set
\begin{equation}
q^\gamma_u = q^\gamma_c = q^\gamma_t\,,
\end{equation}
\begin{equation}
q^\gamma_d = q^\gamma_s = q^\gamma_b\,.
\end{equation}
Below the charm-quark threshold we have
\begin{eqnarray}
n_f=3\,: \quad\,Q^\gamma = \frac{8}{9} q^\gamma_u
+ \frac{4}{9} q^\gamma_d\quad,
\quad \sum_{i=1}^3 e_i^2 = \frac{2}{3}\quad,\quad
\quad \sum_{i=1}^3 e_i^4 = \frac{2}{9} \,.
\end{eqnarray}
Above the charm-quark threshold and below the bottom-quark threshold
the above quantities are changed into
\begin{eqnarray}
n_f=4\,:\quad\,Q^\gamma = \frac{16}{9} q^\gamma_u
+ \frac{4}{9} q^\gamma_d\quad,
\quad \sum_{i=1}^4 e_i^2 = \frac{10}{9}\quad,\quad
\quad \sum_{i=1}^4 e_i^4 = \frac{34}{81} \,.
\end{eqnarray}
Finally above the bottom-quark threshold they become
\begin{eqnarray}
n_f=5\,:\quad\,Q^\gamma = \frac{16}{9} q^\gamma_u
+ \frac{2}{3} q^\gamma_d\quad,
\quad \sum_{i=1}^5 e_i^2 = \frac{11}{9}\quad,\quad
\quad \sum_{i=1}^5 e_i^4 = \frac{35}{81} \,.
\end{eqnarray}
The coefficient functions originate from the following
parton subprocesses. In the Born approximation we have the reaction (fig.2)
\begin{equation}
\gamma^*(q) + \gamma(k) \rightarrow q +\bar q\,,
\end{equation}
where $q$ $(\bar q$) stand for light as well as heavy (anti)-quarks.
The $O(\alpha_s)$ pQCD corrections are given by the one-loop contributions
to process (2.14) (see fig.3) and the gluon bremsstrahlung process
(see fig.4)
\begin{equation}
\gamma^*(q) + \gamma(k) \rightarrow q +\bar q +g \,.
\end{equation}
The parton cross section for the Born reaction (2.14) can be found
in \cite{gr}, \cite{bb} (light quarks) and \cite{dg},
\cite{gr1} (heavy quarks). Notice
that the above reactions are very similar to the ones where the
on-mass-shell photon $\gamma(k)$ is replaced by a
gluon $g(k)$. The cross sections of the photon-induced processes
constitute the Abelian parts of the expressions obtained for the
gluon-induced processes which are presented up to
order $\alpha_s^2$ for the case of massless quarks
in \cite{zn} and in the case of massive quarks in \cite{lrsn1}.
By equating some color factors equal to unity or zero
in the latter expressions one
automatically obtains the cross sections for the
photon-induced processes above.  In the case of massless
quarks the parton cross sections for (2.14), (2.15) contain collinear
divergences which can be attributed to the initial photon being
on-mass-shell. These singularities are removed by mass
factorization in the following way. We define
\begin{equation}
\hat{\cal F}_{k,\gamma}(z,Q^2,\epsilon) = \sum_i\int^1_0\,dz_1
\int^1_0\,dz_2 \delta(z-z_1z_2) \Gamma_{i\gamma}(z_1,M^2,\epsilon)
\,{\cal C}_{k,i}(z_2,\frac{Q^2}{M^2})\,,
\end{equation}
where $\hat{\cal F}_{k,\gamma}(z,Q^2,\epsilon)$ is the parton
structure function, which
is related to the parton cross section in the same way as the
photon structure function $F^\gamma_k(x,Q^2)$ is related to the
cross section $d^2\sigma/dxdy$ in (2.3).
It contains the collinear divergences represented by the parameter
$\epsilon = n-4$ (we use dimensional regularization).
These divergences are absorbed in the transition
functions $\Gamma_{i\gamma}$ $(i=\gamma\,,\,q\,,\,g$) which depend both on
$\epsilon$ and on the mass factorization scale $M$.
The coefficient functions ${\cal C}_{k,i}$ $(i=\gamma\,,\, q\,,\, g$)
are computed in the $\overline{\rm MS}$ scheme
and they appear in the expressions
for the photon structure functions in (2.6). In the case $i=\gamma$,
where the photon is pointlike, the corresponding transition function
is given by
\begin{equation}
\Gamma_{\gamma\gamma} (z,M^2,\epsilon) = \delta(1-z)\,.
\end{equation}
The other transition functions $\Gamma_{i\gamma}$ $(i=q\,,\,g)$ can be
inferred from the Abelian parts of
$\Gamma_{ig}$ \cite{gr}\,,\,\cite{bb}\,,\,\cite{fp}.
In the case of massive quarks
(heavy-flavor production) the parton structure functions do not have mass
singularities. They automatically belong to the
pointlike photon contribution and can be identified with the
coefficient functions ${\cal C}_{k,\gamma}^H$.

The coefficient functions can be expanded in $\alpha_s$ as follows
\begin{eqnarray}
{\cal C}_{k,i} = \,{\cal C}_{k,i}^{(0)} + \frac{\alpha_s(M^2)}{4\pi}
\,{\cal C}_{k,i}^{(1)}+ ...\,,
\end{eqnarray}
with $i=q\,,\,g\,,\,\gamma$ and
\begin{eqnarray}
{\cal C}_{k,\gamma}^H = \,{\cal C}_{k,\gamma}^{H,(0)}
+ \frac{\alpha_s(M^2)}{4\pi}
\,{\cal C}_{k,\gamma}^{H,(1)}+ ...\,.
\end{eqnarray}

In zeroth order of $\alpha_s$ the hadronic coefficient functions are
\begin{eqnarray}
{\cal C}_{2,q}^{(0)}(z,\frac{Q^2}{M^2})= \delta(1-z)\,,
\end{eqnarray}
\begin{eqnarray}
{\cal C}_{L,q}^{(0)}(z,\frac{Q^2}{M^2})= 0\,,
\end{eqnarray}
\begin{eqnarray}
{\cal C}_{k,g}^{(0)}(z,\frac{Q^2}{M^2})= 0\,, \qquad {\mbox (k=2,L)}\,.
\end{eqnarray}
In order $\alpha_S$ the hadronic coefficient functions are given by
\begin{eqnarray}
{\cal C}_{2,q}^{(1)}(z,\frac{Q^2}{M^2})&=& C_F \Big[ \Big\{
\Big(\frac{4}{1-z}\Big)_+ -2 - 2z\Big\} \nonumber \\ &&
\times \Big\{ \ln\frac{Q^2}{M^2} +\ln(1-z) -\frac{3}{4} \Big\}
-2\frac{1+z^2}{1-z} \ln z +\frac{9}{2} +\frac{5}{2} z\nonumber \\ &&
+\delta(1-z)\Big\{3 \ln\frac{Q^2}{M^2}
-9-4\zeta{(2)}\Big\} \Big]\,,
\end{eqnarray}
and
\begin{eqnarray}
{\cal C}_{L,q}^{(1)}(z,\frac{Q^2}{M^2})= C_F \Big[ 4z \Big]\,.
\end{eqnarray}
The above coefficient functions, which emerge via mass factorization
from processess (2.14) and (2.15), (see figs.3 and 4), can be also
inferred from the parton subprocesses
$\gamma^* + q \rightarrow q $ (with one-loop corrections)
and  $\gamma^* + q \rightarrow q + g $ in deep-inelastic
lepton-hadron scattering.
The gluonic coefficient functions ${\cal C}^{(1)}_{k,g}$ although of
order $\alpha_s$ emerge from the $O(\alpha\alpha_s^2)$ process
$\gamma^*(q) + \gamma(k) \rightarrow q + \bar q + q + \bar q$.
Nevertheless they contribute in order $\alpha_s$
after mass factorization has been performed where the corresponding
transition function $\Gamma_{g\gamma}$ leads to the scale dependence
of the gluon density $g^\gamma$.
The gluonic coefficient functions are given by
\begin{eqnarray}
{\cal C}_{2,g}^{(1)}(z,\frac{Q^2}{M^2})&=&n_f T_f \Big[
4\{z^2 +(1-z)^2\}( \ln \frac{Q^2}{M^2} + \ln(1-z) - \ln z)
\nonumber \\ && \qquad\qquad
+32 z (1-z) - 4 \Big]\,,
\end{eqnarray}
\begin{eqnarray}
{\cal C}_{L,g}^{(1)}(z,\frac{Q^2}{M^2})= n_f T_f \Big[ 16 z (1-z) \Big]\,.
\end{eqnarray}
Notice that the latter coefficient functions can be inferred from
the parton subprocess $\gamma^* + g \rightarrow q + \bar q$ in
deep-inelastic lepton-hadron scattering.
The color factors appearing in eqs.(2.23)-(2.26) are
given by $C_F=4/3$ and $T_f = 1/2$ for the case of $SU(3)$.

The photonic coefficient functions in zeroth order of $\alpha_S$ for
massless quarks, denoted by
${\cal C}^{(0)}_{k,\gamma}(z,Q^2/M^2)$,
originate from the Born reaction (2.14) (see (fig.2)). They
can be derived from (2.25), (2.26) as follows
\begin{eqnarray}
{\cal C}_{k,\gamma}^{(0)}(z,\frac{Q^2}{M^2})= \frac{1}{n_fT_f}
\,{\cal C}^{(1)}_{k,g}(z, \frac{Q^2}{M^2}) \,,\quad {\mbox (k=2,L)}\,.
\end{eqnarray}
{}From the same reaction we also obtain
the heavy-flavor contributions which read
\begin{eqnarray}
{\cal C}_{2,\gamma}^{H,(0)}(z,Q^2,m^2)
&=&
\Big[\Big\{ 4 - 8z(1-z) + \frac{16m^2}{Q^2} z (1-3z)
-  \frac{32m^4}{Q^4} z^2 \Big\} L
\nonumber \\ &&
+ \Big\{ -4 +32z(1-z) - 16\frac{m^2}{Q^2} z(1-z)\Big\}
 \sqrt{1-  \frac{4m^2}{s}}
\Big]\,,\nonumber \\
\end{eqnarray}
and
\begin{eqnarray}
{\cal C}_{L,\gamma}^{H,(0)}(z,Q^2,m^2)
&=&
16z(1-z) \Big[ \sqrt{1-\frac{4m^2}{s}} - 2 \frac{m^2}{s} L\Big]\,,
\end{eqnarray}
where $m$ is the heavy-flavor mass
and $\sqrt{s}$ is the c.m. energy
of the virtual photon-real photon system.
Furthermore we have
\begin{equation}
s= (1-z)\frac{Q^2}{z} \quad , \quad
L = \ln\left[\frac{ 1 + \sqrt{1-4m^2/s}}{ 1 - \sqrt{1-4m^2/s}}
\right]\,.
\end{equation}
These formulae were first derived in \cite{ew1},\cite{mg}.

In the next order in $\alpha_s$ process (2.15) (fig.4)
and the one-loop corrections to process (2.14) (fig.3)
give rise to the coefficient functions
${\cal C}_{k,\gamma}^{(1)}(z,{Q^2}/{M^2})$ and
${\cal C}_{k,\gamma}^{H,(1)}(z,Q^2,m^2)$.
In $O(\alpha_s)$ the photonic coefficient functions
for massless quarks
${\cal C}_{k,\gamma}^{(1)}(z,{Q^2}/{M^2})$
can be obtained from the Abelian parts of the
$O(\alpha_S^2)$ contributions to the coefficient functions
${\cal C}_{k,g}(z,{Q^2}/{M^2})$ in \cite{zn} by applying the
same relations as in (2.27).
The corresponding heavy-quark coefficient functions
${\cal C}_{k,\gamma}^{H,(0)}(z,Q^2,m^2)$ can be inferred from the
Abelian parts of the $O(\alpha_S^2)$ contribution
to ${\cal C}_{k,g}(z,Q^2,m^2)$ computed
for heavy-flavor production in \cite{lrsn1}.
Both expressions are too long to be put in this paper.
\footnote{These functions are available
from smith@elsebeth.physics.sunysb.edu.}
We translate the notation used in this paper into those used
in \cite{gr}, \cite{gr1} and \cite{bb} in Table 1,
where we also list the new coefficient functions which were not
used in the earlier calculations in the literature.

\newpage
\mysection{Results}
In this section we will first discuss the $O(\alpha_s)$ corrections
to the hadronic (first) part and the corrections to the
pointlike (second) part of the photon structure functions (2.7).
In particular we focus
our attention on the heavy-flavor contribution (mainly charm)
which enters via the photonic part.
In the literature attempts have been made to implement the
higher-order QCD corrections in the
photon structure function \cite{bb}\,,\cite{gr}.
As we have already pointed out above (2.6) we follow the
prescription in \cite{gr1} which is the same as normally given
for the hadronic structure functions in deep-inelastic lepton-nucleon
scattering \cite{zn}. In this case all the
nonperturbative effects are hidden in the $x$-dependence
of the parton densities $Q^\gamma(x,M^2)$ and $g^\gamma(x,M^2)$ (2.7).
The perturbative parts are given by the
splitting functions and the coefficient functions.
The former appear in the Altarelli-Parisi (AP)
equations which determine the $M^2$-dependence of
the parton densities.
Various parameterizations of the parton densities are given in the
literature (\cite{dg} - \cite{gs}).
However they are all of the leading logarithmic
(LL) type and a consistent $Q^2$ evolution of $F^\gamma_k(x,Q^2)$ can
only be given when the densities are combined with the lowest order
coefficient functions. In our case the latter are given by
${\cal C}^{(0)}_{2,i}$,
${\cal C}^{(1)}_{L,i}$, $(i=q\,,\,g)$ and
${\cal C}^{(0)}_{L,\gamma}$.
Inclusion of the $O(\alpha_s)$ corrections to
${\cal C}_{2,i}$ $(i=q\,,\,g)$ and the contributions
${\cal C}^{(0)}_{2,\gamma}$, ${\cal C}^{(1)}_{L,\gamma}$
requires a next-to-leading-logarithmic (NLL)
parameterization of the parton densities which are available in [8].
If we also want to include ${\cal C}^{(1)}_{2,\gamma}$
one even needs the $O(\alpha\alpha_s^2)$ and $O(\alpha_s^3)$
corrected AP splitting functions
which have not been calculated in the literature. These would
yield the next-to-next-to-leading-logarithmic
approximation (NNLL).
Notice that this problem does not exist for the heavy-flavor
contributions represented by ${\cal C}^{H,(l)}_{k,\gamma}$ $(l=0,1)$
because the latter could be calculated without carrying out
mass factorization, demonstrating that they are independent
of the scale $M^2$.
In spite of the fact that a NLL parametrizaton exists we will
restrict our attention to the LL approximation, because it is
sufficient for our purpose here and because the relatively
poor quality of the data cannot distinguish between the
LL and NLL parametrizations.
Since we omit the NLL and the unknown
 NNLL parton densities the higher-order
pQCD corrections to the coefficient functions have to be considered
as an estimate of how the LL approximation to the
photon structure function will be altered by including
higher-order pQCD effects.

At this moment the LL parton densities give a good
description of the data obtained for the structure function
$F_2^\gamma(x,Q^2)$ over a wide range of $Q^2$ values (see below).
Inclusion of the higher-order QCD corrections leads to a modification
of the nonperturbative parameters describing the
$x$-dependence of the parton densities. Another effect
is that the $Q^2$-dependence of the photon structure function will
be altered when higher-order corrections are included particularly
at large $Q^2$-values.
However the analysis in \cite{gr1} reveals that the addition
of the $O(\alpha_s)$ corrections
${\cal C}^{(1)}_{2,i}$ $(i=q,g)$,
the pointlike photon contributions ${\cal C}^{(0)}_{2,\gamma}$
and the two-loop AP splitting functions to $F_2^\gamma$ hardly changes the
$Q^2$-evolution in the region $5.9 \,\,({\rm GeV}/c)^2 < Q^2 <
110 \,\,({\rm GeV}/c)^2$  accessible to past and present experiments.
This led to the conclusion \cite{gr1} that the LL parameterization for the
parton densities is quite adequate to describe the existing data.
In our opinion this analysis has two drawbacks, which can be
summarized as follows. In spite of the fact that the LL as well as the
NLL parton densities may give a good description
of $F_2^\gamma$ it does not mean that they will provide
us with the same good description for $F_L^\gamma$ since the coefficient
functions for these two structure functions in (2.7) are different.
The same conclusion holds for other photon induced processes like e.g.
photoproduction of heavy flavors \cite{sn}. Another objection is that in
the determination of the LL parton densities the heavy-flavor
contribution which shows up via ${\cal C}^H_{k,\gamma}$ in (2.6),
is neglected. This might be correct for low $Q^2$ in view of the
limited statistics of the available data but is
certainly incorrect for large $Q^2$ as we will see later on.

Besides the theoretical uncertainties one also has to deal with the
quality of the experimental data.
At this moment only data for $F^\gamma_2(x,Q^2)$ are known because
of the experimental limitation $xy^2 << 1$.
The available data have been obtained from various experiments
where $0.03 < x < 0.8$ and
$1.31 < Q^2 < 390$ $({\rm GeV}/c)^2$ \cite{pluto} - \cite{topaz}.
However there exists some hope that at LEP $F^\gamma_L(x,Q^2)$ can
also be measured \cite{ali}.

In the subsequent part of this paper we first discuss how the
LL description for $F_2^\gamma$ is modified by the following
corrections. They are given by:
\begin{enumerate}
\item[I]. The $O(\alpha_s)$ contributions
to the hadronic coefficient functions given by
${\cal C}^{(1)}_{2,i}$ $(i=q,g)$.

\item[II].  The photonic coefficient functions due to light quarks
${\cal C}^{(l)}_{2,\gamma}$ $(l=0,1)$.

\item[III].  The photonic coefficient functions due
to heavy-flavor contributions (mainly charm) represented by
${\cal C}^{H,(l)}_{2,\gamma}$ $(l=0,1)$.
\end{enumerate}
In our calculations we will adopt the two-loop corrected
running coupling constant as presented in Eq.(10) of \cite{admn}.
Further we choose $n_f=4$ in the running coupling constant,
which implies $\Lambda_{\overline{\rm MS}}=0.26$ GeV$/c$.
For the LL parton densities we take
the DG parameterization given in \cite{dg}, where $n_f=3$ (see (2.11)).
The latter is also used in the expressions in Section 2.
The factorization scale is $M = \sqrt{Q^2}$.

In fig.5 we plot $F_2^\gamma(x,Q^2)/\alpha$ for $Q^2= 5.9$
$({\rm GeV}/c)^2$ as a function of $x$, where we compare our results
with the data of PLUTO \cite{pluto}.
The curves represent the following contributions. The solid
line is given by the LL-contribution to $F_2^\gamma$ which
we denote by $F_2^\gamma(LL)$. The dashed line originates from
$F_2^\gamma(LL)$ plus the $O(\alpha_s)$ contributions due to I above.
The latter constitutes the hadronic part of the structure function,
which we will define as ${\rm hadronic} = F^\gamma_2(LL) + O(\alpha_s)$.
If we also include the pointlike photon
part due to II, called light, we get the
dotted line indicated by ${\rm hadronic} + {\rm light}$.
Finally we add the pointlike photon part coming from the
charmed quark (called ${\rm heavy}$) corrected up to
$O(\alpha_s)$ due to III so that the total result
represented by the dotted-dashed line is given by
${\rm hadronic} + {\rm light} + {\rm heavy}$.
The figure reveals an appreciable deviation from the LL prescription
when we include higher-order corrections, in particular in the
higher $x$-region. The main changes are introduced by the
hadronic contribution due to ${\cal C}_{2,q}^{(1)}$
and the pointlike photon contribution due to ${\cal C}_{2,\gamma}^{(0)}$.
The steep rise of the dashed line near
$x=1$ can be attributed to soft-gluon radiation which leads to
the logarithmic terms $(\ln^l(1-z)/(1-z))_+$ $ (l=0,1)$
in ${\cal C}^{(1)}_{2,q}$ (2.23).
The logarithmic term $\ln(1-z)$ occuring in ${\cal C}^{(0)}_{2,\gamma}$
(2.27) (see also (2.25))
is responsible for the large negative correction near $x=1$
and even causes $F_2^\gamma$ to become negative in that region.
The other contributions are unobservable at least within the
errors in the present data.
The gluonic part due to ${\cal C}_{2,g}^{(1)}$ (2.25)
is negligible and the charm contribution ${\cal C}^H_{2,\gamma}$
is small even if one includes
the $O(\alpha_s)$ corrections.
Finally it turns out (see fig.6) that, for $x>0.5$, the
negative contribution due to
${\cal C}^{(0)}_{2,\gamma}$
is partially compensated by adding the $O(\alpha_s)$
corrections represented by ${\cal C}_{2,\gamma}^{(1)}$.
In fig.7 we have compared the charm contribution to
$F_2^\gamma$ with $F^\gamma_2(LL)$. We see that in the region
$0 < x < 0.4$, where charm contributes, it only constitutes about
$20\%$ of $F_2^\gamma(LL)$. If one includes the $O(\alpha_s)$ correction
this becomes much larger and makes up approximately
$30\%$ of $F_2^\gamma(LL)$.

In figs.8-10 we have made the same plots as in
figs.5-7 but now for $Q^2=51$ $({\rm GeV}/c)^2$, where there is
data from the AMY experiment {\cite{amy}.
The figures reveal that the light-quark contribution
becomes smaller (fig.9) whereas the charm contribution
(fig.10) becomes larger with respect to the hadronic part of $F_2^\gamma
$ when $Q^2$ increases.  At $Q^2 = 51$ $({\rm GeV}/c)^2$
the charm contribution constitutes $50\%$
of $F_2^\gamma(LL)$ which is hardly altered when one includes the
$O(\alpha_s)$ correction. The charm component is also larger
than the light-quark part by about a factor of two (compare fig.9
with fig.10).
If one could measure the charm
component cleanly it would provide a good test of pQCD since one
does not need to perform mass factorization to calculate the
corrections to this contribution.

We have also computed the bottom-quark contribution
but it turns out that this is negligible at this $Q^2$ so that we
did not plot it in the figures.
The origin of the suppression of the bottom-quark component
can be attributed to the mass as well as the charge of the bottom quark.
The former is three times larger
than the mass of the charmed quark so there is a phase space suppression.
Moreover $e_H^4$ for the bottom quark contributes
a factor of $1/16$ when compared with the corresponding
$e_H^4$ factor for the charmed quark.
We also studied the plots for $Q^2 = 100$ $({\rm GeV}/c)^2$,
however they did not provide us with additional
useful information when they are compared with those obtained
for $Q^2 = 51$ $({\rm GeV}/c)^2$.

Summarizing the above results we conclude that the light-quark
and heavy-quark
contributions to the pointlike photon part change appreciably
the leading logarithmic (LL) description of the photon structure
function $F_2^\gamma$. This means that the nonperturbative parameters
appearing in the existing parton densities like
DG in \cite{dg} have to be refitted in order to bring the
theoretical description of $F_2^\gamma$ (2.7) in
agreement with the data. Notice that besides the coefficient
functions the authors in ref. \cite{grv} have included the higher-order
AP splitting functions into this analysis.

Bearing in mind the uncertainties above concerning the
leading logarithmic parameterizations of the parton densities
we will give a prediction for the longitudinal
structure function $F_L^\gamma$. For $Q^2 = 5.9$ $({\rm GeV}/c)^2$
we have plotted the hadronic and pointlike photon
parts of $F_L^\gamma$ in fig.11 where the pointlike
photon part is again split up in its light-and
heavy-quark (charm) components. Notice that up to
order $\alpha_s$ the LL description of $F_L^\gamma$
coincides with the $O(\alpha_s)$ corrected hadronic part
since the longitudinal coefficient functions ${\cal C}_{L,i}$
$(i=q\,,\, g)$ are only calculated up to this order.
A glance at fig.11 shows that the hadronic part is heavily
suppressed with respect to the light-quark contribution
to the pointlike photon part, which is due to ${\cal C}^{(0)}_{L,\gamma}$
(2.27).  The latter will be slightly reduced in the region $x > 0.3$
when one includes the $O(\alpha_s)$ correction
${\cal C}_{L,\gamma}^{(1)}$ (fig.12).
The charm contribution is appreciable in the region $x<0.3$ in
particular when one includes the $O(\alpha_s)$
correction (fig.13). The latter increases the Born approximation
for charm production by a factor of two.
The reason the hadronic part is suppressed for $F^\gamma_2$
but not for $F_L^\gamma$ can be traced back to the differences
in the coefficient functions ${\cal C}_{2,q}$
and ${\cal C}_{L,q}$. The former starts already
in order $\alpha_s^0$ whereas the latter starts in
order $\alpha_s$. This has to be compared with the photon
coefficient functions ${\cal C}^{(0)}_{k,\gamma}$
$(k=2,L)$, which are both of order $\alpha_s^0$, so that
${\cal C}^{(1)}_{L,q}$ is suppressed by order $\alpha_s$ with respect
to the other coefficient functions. The second reason is that
${\cal C}_{2,q}^{(1)}$ gets large contributions from
soft-gluon radiation which are absent in ${\cal C}^{(1)}_{L,q}$.
Notice that the gluon density is less important in the
region $0.1 \le x \le 1$.
Moreover it is convoluted with the coefficient functions
${\cal C}_{2,g}^{(1)}$ (2.25) and
${\cal C}_{L,g}^{(1)}$ (2.26) which are both of order $\alpha_s$
and do not contain any soft-gluon enhancements.

If we study $F_L^\gamma$ at $Q^2=51$ $({\rm GeV}/c)^2$
(see fig.14) we observe
that the ratio between the light-quark contribution and the
hadronic component is unaltered with respect to $Q^2=5.9$
$({\rm GeV}/c)^2$ (see also fig.15).
However at $Q^2=51$ $({\rm GeV}/c)^2$ the charm contribution is of the
same size as the light-quark part and becomes much larger
than the hadronic component of $F_L^\gamma$ (fig.16).
The reason that the charm and light-quark contributions
have the same magnitude can be attributed to the fact that
for $Q^2 >> m_c^2$ the coefficient functions
${\cal C}_{L,\gamma}^{(0)}$ (2.27) and
${\cal C}_{L,\gamma}^{H,(0)}$ (2.29) become equal.
{}From fig.16 we also infer that the charm component
is increased by about $30\%$ when the
$O(\alpha_s)$ contribution
${\cal C}_{L,\gamma}^{(1)}$ is included. We also computed the
bottom-quark contribution to $F_L^\gamma$. Like in the case of $F_2^\gamma$
it turned out that the bottom-quark component is negligible so that
it is not shown in the figures. Likewise we did not show
the figures for $Q^2 = 100$ $({\rm GeV}/c)^2$
since they were not qualitatively
different from the ones for $Q^2= 51$ $({\rm GeV}/c)^2$.

A comparison between the plots made for $F_2^\gamma$ and
$F_L^\gamma$ reveals that the hadronic part of $F_L^\gamma$
is heavily suppressed with respect to the pointlike
photon part contrary to what is observed for $F_2^\gamma$.
We do not expect that this feature will be altered if
we had used the NLL parton densities, which are available
in \cite{grv}.
Therefore the longitudinal structure function
$F_L^\gamma$ provides us with a much better test of
pQCD than $F_2^\gamma$. Originally this feature was
expected for $F_2^\gamma$ \cite{ew}. Unfortunately
this expectation did not materialize because of the large hadronic
component in $F_2^\gamma$.
The problem is now left to the experimentalists who
have to try to extract the longitudinal structure
function from the data via the cross section in (2.3).

Summarizing our findings we have seen that the pointlike photon contribution
leads to an appreciable correction to the leading logarithmic
description of $F_2^\gamma$, in particular
in the large $x$ region. Furthermore the pointlike
photon component dominates the longitudinal structure function
$F_L^\gamma$ and overwhelms the hadronic part completely.
The charm-quark contribution, which is relatively small
compared with the light-quark contribution at small $Q^2$, becomes
of the same magnitude as the latter as
$Q^2$ increases (see the plots for $Q^2=51$ $({\rm GeV}/c)^2$).
This feature is characteristic for both structure functions.
Therefore the measurement of the charm contribution
alone would provide a clean test of pQCD.
As far as the $O(\alpha\alpha_s)$ contributions to
the pointlike photon part are concerned we observe that they are
appreciable at small $Q^2$ but become less important relative
to the $O(\alpha)$ contributions
when $Q^2$ becomes larger. This statement holds for both the
light-quark and the heavy-quark contributions.

\vskip 0.5cm

{\bf Acknowledgements}

The work in this paper was supported in part under the
contracts NSF 92-11367 and DOE DE-AC02-76CH03000.
Financial support was also provided by the Texas National
Research Laboratory Commission.

\vfill
\newpage
\newcommand{\caa}{{\cal C}_{2,q}^{(1)} \,\, (2.23)}
\newcommand{\cb}{{\cal C}_{2,g}^{(1)} \,\, (2.25)}
\newcommand{\cc}{{\cal C}_{2,\gamma}^{(0)} \,\, (2.27)}
\newcommand{\cd}{\frac{3\alpha_s}{4\pi}(\frac{2}{3})^4
{\cal C}_{2,\gamma}^{H,(0)} \,\, (2.28)}
\newcommand{\ce}{{\cal C}_{L,q}^{(1)} \,\, (2.24)}
\newcommand{\cf}{{\cal C}_{L,g}^{(1)} \,\, (2.26)}
\newcommand{\cg}{{\cal C}_{L,\gamma}^{(0)} \,\, (2.27)}
\newcommand{\ch}{{\cal C}_{L,\gamma}^{H,(0)} \,\, (2.29)}
\newcommand{\ci}{{\cal C}_{k,\gamma}^{(1)}\,\,\cite{zn}}
\newcommand{\cj}{{\cal C}_{k,\gamma}^{H,(1)} \,\, \cite{lrsn1}}
\newcommand{\ba}{B^{(n)}_{\rm NS}, B^{(n)}_\psi \,\, (4.10)}
\newcommand{\bb}{B^{(n)}_{\rm G} \,\, (4.11)}
\newcommand{\bc}{B^{(n)}_{\gamma} \,\, (4.12)}
\newcommand{\be}{B_{\rm NS}, B_q \,\, (3.7)}
\newcommand{\bff}{B_{\rm G} \,\, (3.7)}
\newcommand{\bg}{B_{\gamma} \,\, (3.7)}
\newcommand{\bh}{\frac{1}{x}F^\gamma_{2,c} \,\, (2.13)^*}

\centerline{\bf \large{Table 1.}}
\vspace{2cm}

\hspace{-0.1cm}\begin{tabular}{||c||c|c||} \hline
{\rm this paper} & \cite{bb} & \cite{gr}\,,\,\cite{gr1}${}^*$ \\ \hline

$\caa $ & $\ba$ & $\be$ \\  \hline
$\cb $ & $\bb$ & $\bff$ \\  \hline
$\cc $ & $\bc$ & $\bg$ \\  \hline
$\cd $ &  ---  & $\bh$ \\  \hline
$\ce $ &  ---  &  ---  \\  \hline
$\cf $ &  ---  &  ---  \\  \hline
$\cg $ &  ---  &  ---  \\  \hline
$\ch $ &  ---  &  ---  \\  \hline
$\ci $ &  ---  &  ---  \\  \hline
$\cj $ &  ---  &  ---  \\  \hline
\end{tabular}

\vspace{0.5cm}

Notations in several papers for the hadronic and photonic
coefficient functions. Notice that the expressions in
\cite{bb} are in Mellin transform space. The blanks mean
that these contributions were not considered in the papers quoted.
%

\end{document}